\documentclass[12pt]{article}
\usepackage{indentfirst}
\usepackage{graphicx}
\usepackage{placeins}
\usepackage{slashed}
\usepackage{gensymb}
\usepackage{amsmath}
\usepackage{amssymb}
\usepackage{feynmp-auto}
\usepackage{multirow}
\usepackage[hidelinks]{hyperref}
\usepackage{sourceserifpro}
\usepackage[T1]{fontenc}
\usepackage{cite}

\newcommand\pubnumber{}
\newcommand\pubdate{\today}

\def\napoli{Physikalisches Institut and Bethe Center for Theoretical Physics,\\
Bonn University, 53115 Bonn, Germany}
\def\support{\footnote{zhongyi@th.physik.uni-bonn.de}}

\def\Title#1{\begin{center} {\Large #1 } \end{center}}
\def\Author#1{\begin{center}{ \sc #1} \end{center}}
\def\Address#1{\begin{center}{ \it #1} \end{center}}

\newcommand\pubblock{\rightline{\begin{tabular}{l} \pubnumber\\
         \pubdate  \end{tabular}}}
\newenvironment{Abstract}{\begin{quotation}  }{\end{quotation}}

\def\Acknowledgements{\bigskip  \bigskip \begin{center} \begin{large}
             \bf ACKNOWLEDGEMENTS \end{large}\end{center}}

\begin{document}

\def\gsim{\:\raisebox{-0.5ex}{$\stackrel{\textstyle>}{\sim}$}\:}
\def\lsim{\:\raisebox{-0.5ex}{$\stackrel{\textstyle<}{\sim}$}\:}

\begin{titlepage}
\pubblock

\vfill
\Title{Constraints on $U(1)_{L_\mu-L_\tau}$ from LHC Data}
\Author{Manuel Drees\footnote{drees@th.physik.uni-bonn.de}, Meng
Shi\footnote{mengshi@physik.uni-bonn.de} and Zhongyi Zhang\support}
\Address{\napoli}

\begin{Abstract} 
  In this study, we apply LHC data to constrain the extension of the
  Standard Model by an anomaly--free $U(1)_{L_\mu-L_\tau}$ gauge
  group; this model contains a new gauge boson ($Z^\prime$) and a
  scalar dark matter particle ($\phi_{\rm DM}$). We recast a large
  number of LHC analyses of multi--lepton final states by the ATLAS
  and CMS collaborations. We find that for
  $10 \ {\rm GeV} < m_{Z'} < 60$ GeV the strongest constraint comes
  from a dedicated $Z'$ search in the $4\mu$ final state by the CMS
  collaboration; for larger $Z'$ masses, searches for final states
  with three leptons plus missing $E_T$ are more sensitive. Searches
  for final states with two leptons and missing $E_T$, which are
  sensitive to $Z'$ decays into dark matter particles, can only probe
  regions of parameter space that are excluded by searches in the $3$
  and $4$ lepton channels. The combination of LHC data excludes values
  of $Z'$ mass and coupling constant that can explain the deficit in
  $g_\mu - 2$ for $4 \ {\rm GeV} \leq m_{Z'} \leq 500$ GeV. However,
  for much of this range the LHC bound is weaker than the bound that
  can be derived from searches for ``trident'' events in
  neutrino--nucleus scattering.
\end{Abstract}
\vfill

\end{titlepage}
\def\thefootnote{\fnsymbol{footnote}}
\setcounter{footnote}{0}

\section{Introduction}
\label{sec:intro}

There are several reasons for considering the extension of the gauge
group of the Standard Model (SM) by another Abelian $U(1)$ factor. It
is usually assumed that the new gauge boson couples universally to all
three generations of quarks, in order to avoid constraints from
flavor--changing neutral currents. If we further insist that the gauge
group should be anomaly--free within the SM matter content (possibly
extended by right--handed neutrinos, but without other exotic chiral
fermions), there are only four different possibilities. These can be
written as $B-L$ \cite{Mohapatra:1980qe}, as well as the purely
leptonic $L_e-L_\mu$, $L_\mu-L_\tau$, and $L_e-L_\tau$
\cite{He:1990pn}; of course, linear combinations of these four groups
are also possible. In the $U(1)_{B-L}$ model, which does require
right handed neutrinos, the new gauge boson couples to both quarks and
(charged) leptons. This model is therefore tightly constrained by
searches for di--lepton resonances at hadron colliders, in particular
at the LHC \cite{Aaboud:2017buh, Sirunyan:2018exx}; if the coupling of
the new $U(1)$ is comparable to that of the $U(1)_Y$ of the SM, these
searches exclude $Z'$ masses below several TeV.\footnote{See
  ref.\cite{Amrith:2018yfb} for a very recent assessment on current
    constraints on the $B-L$ model.}

In the other three models the cancellations of anomalies occur
between different generations without the requirements of extra
fermions \cite{He:1990pn}. LEP data strongly constrain the $L_e-L_\mu$
and $L_e-L_\tau$ models. While we are not aware of dedicated analyses
of LEP data in the framework of these models, for $m_{Z'} > 300$ GeV
or so the $Z'$ propagator at LEP energies ($\sqrt{s} \leq 209$ GeV)
can be approximated by a constant, in which case limits on contact
interactions apply. In particular, ALEPH data on
$e^+e^- \rightarrow \mu^+\mu^-$ \cite{Schael:2006wu} imply
$m_{Z'}/g_{e\mu} > 1.1$ TeV for the $L_e-L_\mu$ model, whereas OPAL
data on $e^+e^- \rightarrow \tau^+\tau^-$ \cite{Abbiendi:2003dh} lead
to the bound $m_{Z'} / g_{e\tau} > 0.94$ TeV for the $L_e - L_\tau$
model. For smaller $Z'$ masses, where propagator effects become
important, the bound will be even stronger.

In contrast, the $L_\mu - L_\tau$ model does not predict any new
interaction for the electron. Its gauge boson can therefore only be
produced through higher--order processes in $e^+e^-$ collisions, by
emission off a charged or neutral lepton of the second or third
generation. These final states can also be produced at the LHC
\cite{Ma:2001md}, which has accumulated a far larger number of
di--muon and di--tau events than LEP did. In this paper we therefore
focus on LHC data. Note also that the $U(1)_{L_\mu-L_\tau}$ model can
accommodate successful neutrino masses even with the simplest Higgs
sector \cite{Asai:2017ryy, Asai:2018ocx}, and can be extended to
contain a dark matter particle that is charged under the new symmetry
but easily satisfies the stringent direct search constraints
\cite{Biswas:2016yan, Biswas:2016yjr}. In principle, this model could
also explain the difference between SM prediction and measurement of
the anomalous magnetic moment of the muon $(g_\mu-2)$; however, bounds
on $\nu_\mu N \rightarrow \nu_\mu \mu^+\mu^- N$ ``trident'' production
\cite{Altmannshofer:2014pba}, where $N$ stands for some nucleus,
exclude this possibility for $m_{Z'} > 0.5$ GeV.

The other existing constraint in the $Z^\prime$ mass range relevant
for searches at the LHC comes from analyses of $Z$ decays into four
charged leptons \cite{Rainbolt:2018axw}. In particular,
ref.\cite{Sirunyan:2018nnz} is a CMS analysis constraining this model
using $Z \rightarrow 4\mu$ decays. This search is obviously only
sensitive to relatively light $Z'$, $m_{Z'} < m_Z$. LHC prospects for
this model have been discussed previously \cite{Harigaya:2013twa,
  delAguila:2014soa, Elahi:2015vzh}, with ref.~\cite{Elahi:2015vzh}
focussing on the case $m_{Z'} \leq m_Z/2$; however, these papers did
not attempt to use actual LHC data to constrain the model.

In contrast, we consider a comprehensive set of LHC analyses for final
states with two, three or four charged leptons in the final state,
where a lepton $l$ for us means a muon or a hadronically decaying
$\tau$. Final states with fewer than four charged leptons are also
required to contain some missing transverse momentum
$\slashed{E_T}$. In particular, final states with only two charged
leptons plus $\slashed{E_T}$ are sensitive to $Z'$ decays into dark
matter particles, which also reduce the branching ratios for $Z'$
decays into $\mu$ or $\tau$ pairs. $\tau \rightarrow \mu$ decays
contribute to muonic final states, if typically with reduced
efficiency since the muon produced in $\tau$ decays is obviously
softer than the parent $\tau$. In principle, $\tau \rightarrow e$
decays can also populate final states with electrons. However, the
small branching ratio (about $18\%$) and again reduced efficiency
imply that final states with electrons will not be as sensitive as
those only containing muons or hadronically decaying $\tau$
leptons. We use the {\tt CheckMATE} framework
\cite{Drees:2013wra,Dercks:2016npn}. Only a few of the analyses we
applied had already been included in {\tt CheckMATE}. We included a
total of $281$ new signal regions defined in $28$ different
papers.\footnote{Most of the experimental papers we used also include
  signal regions containing electrons. We did not consider those, for
  the reasons explained above. By current policy an analysis can
  become part of the official {\tt CheckMATE} release only if {\em
    all} of the signal regions defined in this analysis are
  encoded. Our ``private'' version of {\tt CheckMATE} is available
  upon request.} We find that the specialized $Z'$ search
\cite{Sirunyan:2018nnz} based on $4\mu$ final states is indeed most
sensitive for $10 \ {\rm GeV} \leq m_{Z'} \leq 60$ GeV; for larger
masses, analyses of final states containing only three charged leptons
are more sensitive.

The full $SU(3)_c\times SU(2)_L\times U(1)_Y\times U(1)_{L_\mu-L_\tau}$ model
introduced in \cite{Biswas:2016yan, Biswas:2016yjr} contains not only the new
mediator and DM particle, but also an extra Higgs boson to break the new
$U(1)$ as well as SM singlet right--handed neutrinos for a see--saw
generation of realistic neutrino masses. The extra Higgs boson plays a
significant role in the dark matter phenomenology, but it can contribute to
the final states we consider only if its mixing angle with the $SU(2)$
doublet Higgs boson responsible for electroweak symmetry breaking is
relatively large. We ignore this possible source of additional signal events.
The main free parameters are thus the mass of the $Z'$ and the strength of
its coupling to $\mu$ and $\tau$ leptons; the branching ratio for $Z'$ decays
into dark matter particles also plays a (lesser) role.

The reminder of this Letter is organized as follows. In
Sec.~\ref{sec:2}, we briefly describe the parts of the
$U(1)_{L_\mu-L_\tau}$ model \cite{Biswas:2016yan, Biswas:2016yjr} that
are relevant for the LHC searches we consider. The application to LHC
data is discussed in Sec.~3, both for vanishing and non--vanishing
branching ratio for $Z'$ decays into dark matter particles. Finally,
Sec.~\ref{sec:4} contains our summary and conclusions.

\section{The Simplified $U(1)_{L_\mu-L_\tau}$ Model}
\label{sec:2}

The $SU(3)_c\times SU(2)_L\times U(1)_Y\times U(1)_{L_\mu-L_\tau}$
model contains a new gauge boson $Z^\prime$ for the local
$U(1)_{L_\mu-L_\tau}$ symmetry; the corresponding field strength
tensor is
$\mathcal{Z}^\prime_{\mu\nu} \equiv \partial_\mu Z^\prime_\nu
-\partial_\nu Z^\prime_\mu$. As usual, we write its interaction with
other particles using the covariant derivative instead of the normal
partial derivative, i.e.
$\partial_\mu\rightarrow D_\mu = \partial_\mu - i g_{\mu\tau} q
Z^\prime_\mu$, where $g_{\mu\tau}$ is the new gauge coupling and $q$
is the $L_\mu - L_\tau$ charge of the particle in question. The model
also contains a complex scalar $\phi_{\rm DM}$, which is singlet under
the gauge group of the SM but carries $L_\mu - L_\tau$ charge
$q_{\rm DM}$. The new part of the complete Lagrangian that is relevant
for our analysis is thus given by
\begin{eqnarray} \label{lag}
  \mathcal{L}_{\rm new} &=& (D_\mu\phi_{\rm DM})^* D^\mu\phi_{\rm DM} -
  m_{\rm DM}^2 \phi^*_{\rm DM}\phi_{\rm DM}
  - \frac{1}{4} \mathcal{Z}^\prime_{\mu\nu} \mathcal{Z}^{\prime\mu\nu}+
\frac{1}{2}m_{Z^\prime}^2Z^{\prime\mu} Z^\prime_\mu\\ \nonumber  &+&
g_{\mu\tau}(\bar{\mu}\slashed{Z}^\prime \mu + \bar{\nu}_\mu\slashed{Z}^\prime
\nu_\mu - \bar{\tau}\slashed{Z}^\prime \tau - \bar{\nu}_\tau\slashed{Z}^\prime
\nu_\tau). 
\end{eqnarray}

The LHC signals we consider originate from the production and decay of
(nearly) on--shell $Z^\prime$ bosons. At leading order the $Z^\prime$
can only decay into second or third generation leptons, and possibly
into DM particles. The corresponding partial widths are given by
\begin{equation} \label{gam_l}
\Gamma(Z^\prime \rightarrow l^+ l^-) =
	\frac{g_{\mu\tau}^2 m_{Z^\prime}} {12\pi} \sqrt{1-4z_l}(1+2z_l)\,, \ \
{\rm for} \ 	l=\mu,\,\tau; 
\end{equation}
\begin{equation} \label{gam_phi}
\Gamma(Z^\prime\rightarrow \phi_{\rm DM}\phi^*_{\rm DM}) =
\frac{q^2_{\rm DM} g^2_{\mu\tau} m_{Z^\prime}}{48\pi} (1-4z_{\rm DM})^{3/2}\,,
\end{equation} 
where $z_X\equiv \frac{m_X^2}{m_{Z^\prime}^2}$. The partial width for
$Z^\prime$ decays into one flavor ($\mu$ or $\tau$) of neutrino is
half of that given in eq.(\ref{gam_l}), since only the left--handed
neutrinos are light enough to contribute. In our analysis we only
consider scenarios where the total $Z^\prime$ width is smaller than
$m_{Z^\prime}$, since otherwise perturbation theory is not
reliable. This translates into the condition
\begin{equation} \label{eq:Pert}
  q^2_{\rm DM}(1-4z_{\rm DM})^{3/2} + 4 \sum_{l=\mu,\,\tau} \sqrt{1-4z_l}(1+2z_l)
   + 4 < 48 \pi/g^2_{\mu\tau}\,.
\end{equation} 
This bound is always satisfied for $g_{\mu\tau} \leq 3$ and
$q_{\rm DM} \leq 2$.

\section{Application to LHC Data} 
\label{sec:3} 
\setcounter{footnote}{0}

\begin{figure}[htb]
\includegraphics[width=0.33\textwidth]{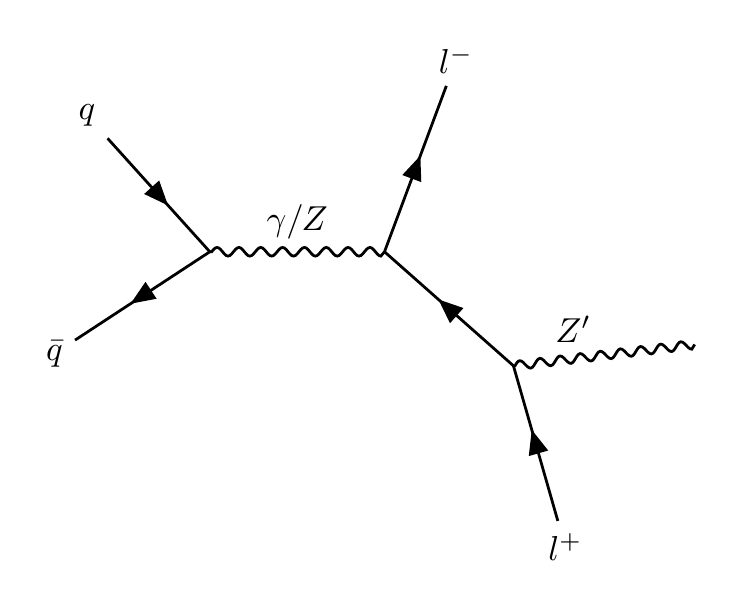}
\includegraphics[width=0.33\textwidth]{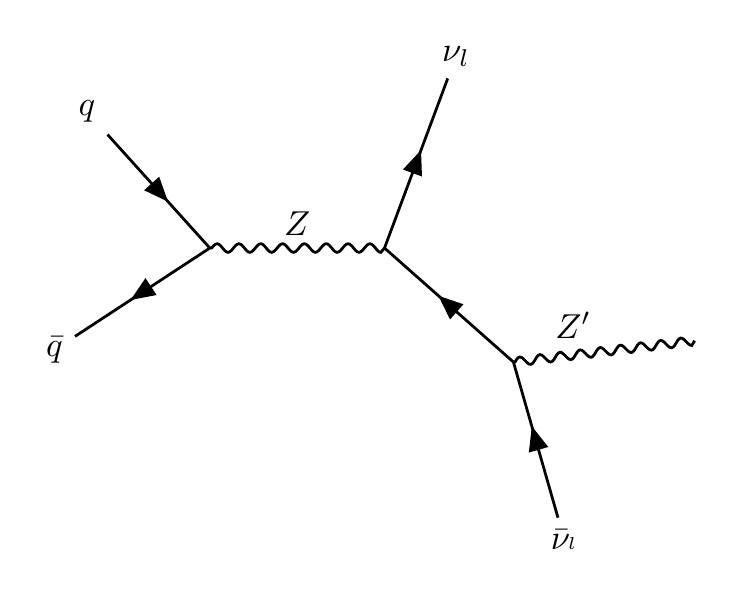}
\includegraphics[width=0.33\textwidth]{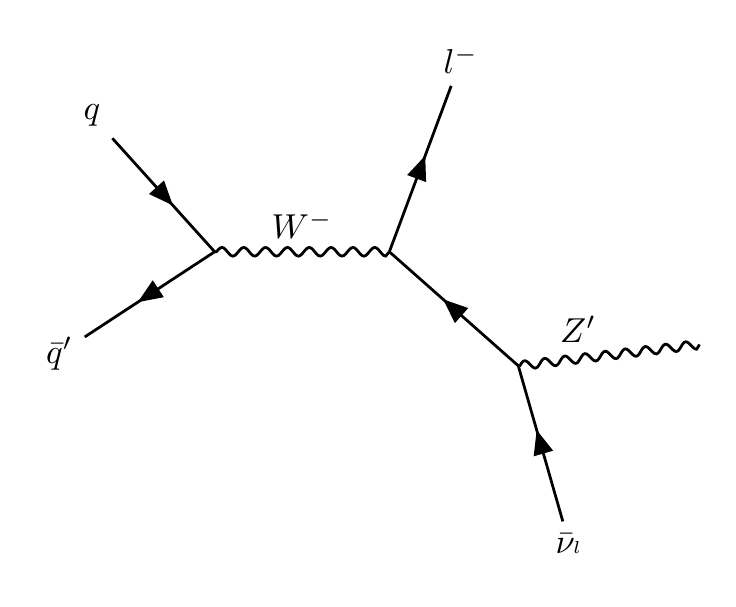}
\caption{Examples of Feynman diagrams for
  $pp\rightarrow Z^\prime l^+ l^-$ (left),
  $pp \rightarrow Z^\prime \nu_l \bar{\nu}_l$ (center) and
  $pp \rightarrow Z^\prime l \nu_l$ (right). For the left diagram,
  both visible (leptonic) and invisible $Z^\prime$ decays (into
  neutrinos or DM particles) contribute to signal processes, but for
  the central and right diagram only $Z^\prime$ decays into a charged
  lepton pair were considered. The $Z^\prime$ boson can also be
  emitted off the other lepton, and $W^+$ exchange diagrams also
  contribute. In the event generation the $Z^\prime$ is allowed to be
  off--shell.}
\label{fig:feyn}
\end{figure}

At tree--level the only SM particles our $Z^\prime$ boson couples to
are leptons of the second and third generation. These can be
pair--produced via neutral or charged current Drell--Yan
processes. The leading--order $Z^\prime$ production processes are
based on these Drell--Yan reactions, with a $Z^\prime$ boson being
emitted off the lepton line, see Fig.~\ref{fig:feyn}.

If the primary Drell--Yan process produces an $l^+l^-$ pair (left diagram),
$Z^\prime \rightarrow l'^+l'^-$ decays lead to final states containing
four charged leptons, where flavor $l'$ may be the same or different from
$l$ (with $l,\, l' \in \{\mu,\tau\}$). Invisible $Z^\prime$ decays, into
neutrinos or DM particles, lead to final states with an opposite--sign
same--flavor charged lepton pair plus missing $E_T$.

If the primary Drell--Yan reaction produces a $\nu_l \bar\nu_l$ pair
(middle diagram), $Z^\prime$ decays into charged leptons again lead to
$l^+l^- \slashed{E_T}$ final states. For this production process
invisible $Z^\prime$ decays do not result in a detectable final
state.\footnote{If a hard parton is emitted off the initial state this
  process would contribute to monojet production; however, it would merely
  be a higher--order correction to monojet production in the SM, and
  would thus certainly not lead to a detectable signal.}

Finally, if the primary Drell--Yan reaction produces a $l^- \bar\nu_l$
pair or its charge conjugate (right diagram), $Z^\prime$ decays into
charged leptons leads to final states of the type
$l^\pm l'^+ l'^- \slashed{E_T}$, where the $l$ and $l'$ may again be
the same or different flavors. In this case invisible $Z^\prime$
decays lead to final states with a single charged lepton plus missing
$E_T$. This can be considered a higher--order correction to the SM
charged--current Drell--Yan reaction, and will certainly have a far
worse sensitivity than the $3l + \slashed{E_T}$ final state.

Of course, experimentally a $\mu$ and a $\tau$ look very different. In
fact, primary muons and muons from tau decays cannot be distinguished
reliably; we will just add these contributions. For reasons described
in the Introduction, we do not consider final states containing
electrons, which might be produced in tau decays. However, we do
consider final states including hadronically decaying tau leptons,
which we denote by $\tau_h$.

Altogether, we thus consider the following distinct final states:
$3\mu$, $4\mu$, $m\mu+n\tau_h$ ($m+n>2,\,n\neq 0$),
$2\tau_h+\slashed{E_T}$, $\mu\tau_h+\slashed{E_T}$, and
$2\mu + \slashed{E_T}$. The corresponding LHC analyses we recast are
summarized in Table~\ref{tab:01}. To that end we used the {\tt
  CheckMATE 2} framework \cite{Dercks:2016npn}, which in turn uses
{\tt Delphes 3}~\cite{deFavereau:2013fsa} to simulate the
CMS~\cite{Chatrchyan:2008aa} and ATLAS~\cite{Aad:2008zzm}
detectors. It should be noted that {\tt CheckMATE} also uses several
other public tools \cite{deFavereau:2013fsa, Cacciari:2011ma,
  Cacciari:2005hq, Cacciari:2008gp, Read:2002hq, Lester:1999tx,
  Barr:2003rg, Cheng:2008hk, Bai:2012gs, Tovey:2008ui,
  Polesello:2009rn, Matchev:2009ad}. As mentioned in the Introduction,
we encoded a total of $281$ new signal regions; we also used a few
searches for superparticles in multi--lepton final states which had
already been included in {\tt CheckMATE}.

\begin{table}[htb]
\begin{center}
\begin{tabular}{|c|c|c|c|}
\hline
List of Analyses & \multicolumn{3}{|c|}{Center--of--mass energy}\\ \hline
Topologies & 7 TeV & 8 TeV & 13 TeV\\ \hline
$2\mu+\slashed{E_T}$ &
\cite{Khachatryan:2014aep, Chatrchyan:2013yaa} &
\cite{Khachatryan:2014aep, Khachatryan:2016vnn, 
Khachatryan:2015sga, Aad:2015cxa} & 
\cite{Sirunyan:2017onm, Sirunyan:2017qaj, Sirunyan:2017qfc,
Sirunyan:2018iwl, Sirunyan:2018ubx, Sirunyan:2018egh,
Sirunyan:2018exx, Sirunyan:2018nwe, Sirunyan:2018lul, 
Aaboud:2018jiw, Aaboud:2018puo}\\ \hline
$(2\tau_h\ or\ \mu\tau_h)+\slashed{E_T}$ & & &
\cite{Sirunyan:2018fpy, Sirunyan:2018vig, Aaboud:2017sjh, 
Aaboud:2018jff}\\ \hline
$3\mu\ or\ 4\mu$ & & \cite{Khachatryan:2015wka} &
\cite{Sirunyan:2017zjc, Sirunyan:2018ubx, Sirunyan:2018shy,
Sirunyan:2018egh, Sirunyan:2018nnz, Aaboud:2017rwm, Aaboud:2017vzb, 
Aaboud:2018jiw, Aaboud:2018puo, Aaboud:2018qcu, Sirunyan:2018mbx, 
Sirunyan:2017lae}\\ \hline
$m\mu+n\tau_h$ & & \multirow{2}{*}{ \cite{Khachatryan:2017mnf} } &
\multirow{2}{*}{ \cite{Sirunyan:2018ubx, Sirunyan:2018shy, 
Sirunyan:2018mbx, Sirunyan:2017lae} }\\
$(m+n>2,\,n\neq 0)$ & & & \\ \hline
\end{tabular}
\caption{All analyses used in this study.}
\label{tab:01}
\end{center}
\end{table}

In order to simulate the signal, we used {\tt
FeynRules}~\cite{Alloul:2013bka} to produce a model file output in {\tt UFO}
format~\cite{Degrande:2011ua}. Parton--level events were generated by {\tt
MadGraph}~\cite{alwall2011madgraph}. Specifically, we defined charged leptons
(meaning $\mu^-$ and $\tau^-$) and invisible particles ($\mu$ and $\tau$
neutrinos or antineutrinos as well as DM particles). The $2l$ signal events
were generated by specifying {\tt MadGraph} events containing a charged
lepton--antilepton pair plus two missing particles; for the $3l$ signal,
${\tt MadGraph}$ generated events with three charged leptons and one missing
particle; and the $4l$ signal started from {\tt MadGraph}--generated events
with two pairs of charged leptons. In all cases only diagrams containing one
$Z^\prime$ propagator (i.e. two new couplings) were generated.

This means that the $Z^\prime$ boson is allowed to be off--shell, but
interference between $Z^\prime$ and $Z$ or photon exchange is not included.
These interference terms formally vanish in the narrow width approximation,
i.e. for $\Gamma_{Z'} \rightarrow 0$. These terms are therefore expected to
be more important for larger coupling $g_{\mu\tau}$, which in turn are
allowed for larger $m_{Z'}$, as discussed quantitatively below. However, we
found that even for the largest coupling we consider, which respects the
perturbativity constraint (\ref{eq:Pert}), the interference contribution to
the cross section after cuts is at most $6\%$ of the squared $Z'$ exchange
contribution. This is considerably less than the effect of typical QCD NLO
corrections, which we also ignore. Note also that in the high mass region
($m_{Z^\prime}>100$ GeV), where the upper limit of $g_{\mu\tau}$ is sizable
and considered offering noticeable interference contribution, we found the
interference terms to be positive, so ignoring them is conservative.

These {\tt MadGraph} events were passed on to {\tt Pythia
  8.2}~\cite{sjostrand2015introduction} for parton showering and
hadronization, and then to {\tt CheckMATE 2}~\cite{Dercks:2016npn}
which applies the selection cuts defined by the designated search
regions and decides whether the given model is excluded by these
searches or not.

We performed separate comparisons to $2l, \ 3l$ and $4l$ searches; we
remind the reader that $l$ here means a muon or a hadronically
decaying $\tau$ lepton. Some of the analyses we apply used data taken
at $\sqrt{s} = 7$ or $8$ TeV, which required separate event
generation. However, at the end the analyses of data taken at
$\sqrt{s} = 13$ TeV, many of which were published quite recently,
always proved more constraining. Moreover, we find that replacing a
muon in the final state by a hadronically decaying $\tau$ always
reduces the sensitivity. The branching ratio for hadronic $\tau$
decays is about $65\%$, but the $\tau-$tagging efficiency is well
below the efficiency of identifying a muon, and QCD jets are much more
likely to be misidentified as a hadronically decaying $\tau$ than as a
muon. Nevertheless $\tau$ leptons do contribute to the final sensitivity,
through $\tau \rightarrow \mu$ decays. 

In the following we will present constraints on the $L_\mu - L_\tau$
gauge boson in two different scenarios. We begin with scenarios where
the $Z^\prime$ boson does not decay into dark matter particles, either
because $q_{\rm DM} = 0$ or because $m_{\rm DM} > m_{Z^\prime}/2$. The
strengths of all signals we consider can then be computed uniquely in
terms of only two parameters: the mass $m_{Z^\prime}$ and the coupling
$g_{\mu\tau}$. We generate at least $20,000$ events for each
combination of $Z^\prime$ mass and coupling; if the total error in the
most relevant signal region is dominated by Monte Carlo statistics, we
generate additional events. Since the signal rates to good
approximation scale like $g^2_{\mu\tau}$, we typically only need to
try three to four values of the coupling in order to determine its
upper bound for a given value of $m_{Z^\prime}$.

\begin{figure}[htb]
\includegraphics[width=0.5\textwidth]{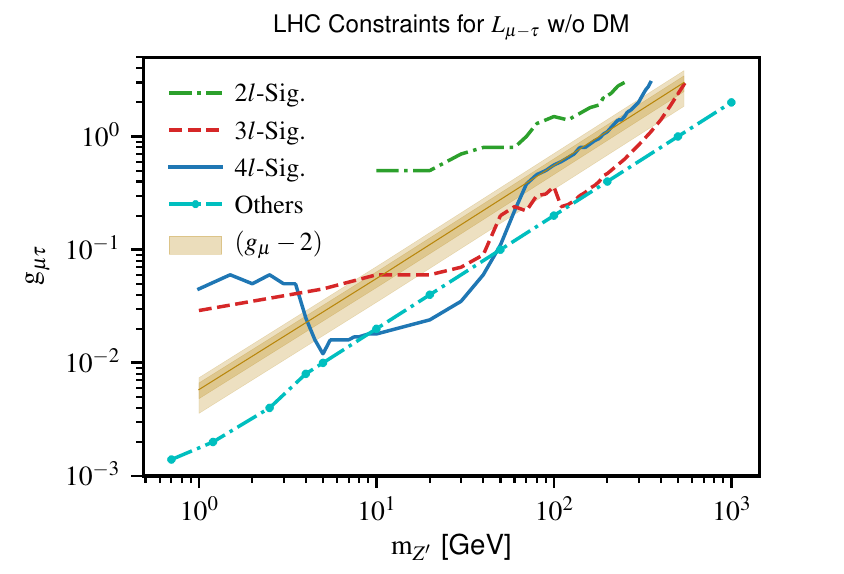}
\includegraphics[width=0.5\textwidth]{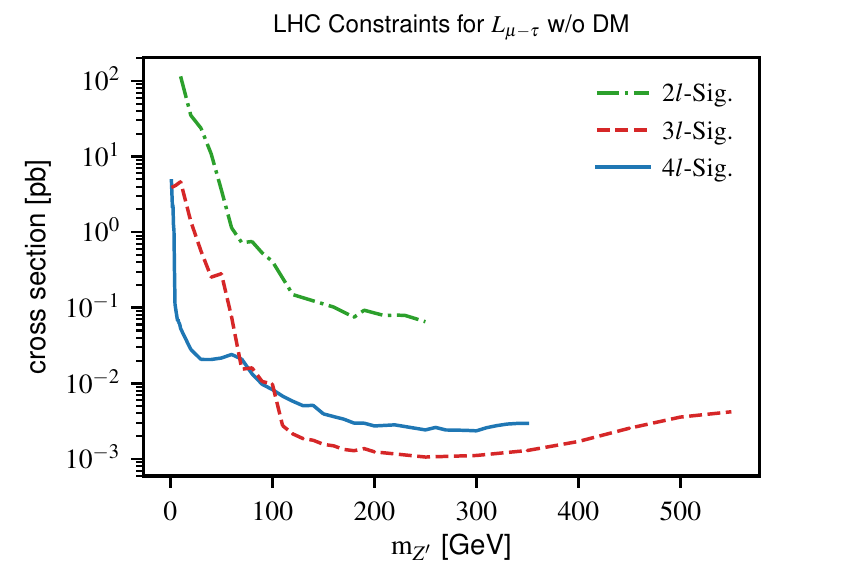}
\caption{The upper limit on the new coupling $g_{\mu\tau}$ (left) and
  the corresponding cross section before cuts (right). The left frame
  also shows the value of the coupling indicated by the measurement of
  the anomalous magnetic moment of the muon (shaded area), as well as a
  summary of existing constraints (lower dot--dashed curve); see the
  text for further details.}
\label{rst:01}
\end{figure}

In the left frame of Fig.~\ref{rst:01} we show upper bounds on
$g_{\mu\tau}$ that have been derived in this manner as functions of
$m_{Z^\prime}$. The figure shows separate bounds from analyses of
final states with two (green dot--dashed curve), three (red dashed
curve) and four (dark blue solid curve) charged leptons. The right
frame shows the upper bounds on the corresponding total cross
sections, which include the branching ratios for $Z^\prime$ decays but
count each $\tau$ as a charged leptons, irrespective of its decay. The
curves terminate in the region of large $Z^\prime$ mass when the
perturbativity limit (\ref{eq:Pert}) is reached. The curves aren't
always smooth. The reason is that {\tt CheckMATE} uses the signal
region with the best {\em expected} sensitivity to set the bounds.
This avoids ``look elsewhere'' effects, but can lead to
discontinuities when the relevant signal region changes. Finally,
we do not show bounds from $2l$ final states for $m_{Z^\prime} < 10$
GeV since the cut efficiency becomes very poor there, i.e. we would
need to generate a very large number of events in order to derive
reliable results; we did not do that since the resulting bound will
surely again be worse than that from $3l$ and $4l$ analyses.

The left frame also shows the value of $g_{\mu\tau}$ where the full
theory prediction, including $Z^\prime$ exchange, reproduces the
measured value of $g_\mu - 2$. The brown solid line corresponds to the
central value, whereas the darker and lighter shaded regions allow too
reproduce $g_\mu-2$ up to $1$ and $2$ standard deviations, respectively.
Here we use
$$\Delta a_\mu=a_\mu^{exp}-a_\mu^{th}=(29.0\pm 9.0)\times 10^{-10}$$
from \cite{Jegerlehner:2009ry}, which is was also used in the
non--collider studies \cite{Biswas:2016yjr, Biswas:2016yan} we
discussed previously.

Finally, the lower dot--dashed line in the left frame summarizes
non--LHC bounds.  For $m_{Z^\prime} > 4$ GeV the results from non--LHC
data come from our interpretation of the CCFR measurement of the cross
section for ``trident'' production \cite{Mishra:1991bv}. We used the
CL$_{\rm S}$ method to set the $95\%$ c.l.  limit, which is also
employed by {\tt CheckMATE}. The resulting bound on $g_{\mu\tau}$ is
$\sim 20\%$ weaker than that derived by taking the central value of
the CCFR cross section plus $1.64$ times the CCFR error as upper bound
on the cross section, which seems to have been done in
\cite{Altmannshofer:2014pba}; note that the cross section measured by
CCFR is somewhat below the SM prediction.\footnote{The CHARM--II
  collaboration also measured this cross section, with a different
  neutrino beam, and found a result somewhat larger than, but
  compatible with, the SM prediction \cite{Geiregat:1990gz}. Naively
  averaging the two measurements of
  $\sigma_{\rm exp} / \sigma_{\rm SM}$ leads to a very similar bound
  on $g_{\mu\tau}$ when using the CL$_{\rm S}$ method.} For
$m_{Z^\prime} < 4$ GeV the best non--LHC bound comes from $4\mu$
searches by the BaBar collaboration \cite{TheBABAR:2016rlg}. We show a
smoothed--out version of the actual bound, which fluctuates rapidly by
$\sim \pm 30\%$ around this line. In \cite{Chun:2018ibr} it was shown
that bounds from tests of lepton universality are always weaker than
that from the neutrino trident experiments in the parameter region we
focus on ($m_{Z^\prime}\leq 500$ GeV). We therefore do not show these
constraints in Fig.~\ref{rst:01}.

As mentioned above, there is only one published analysis of LHC data
that specifically searches for the $L_\mu - L_\tau$ gauge boson
\cite{Sirunyan:2018nnz}; it covers the mass range
$5\ {\rm GeV} <m_{Z^\prime}< 70$ GeV using $Z \rightarrow 4\mu$ decays
in the CMS detector. Our {\tt CheckMATE} based recast of this analysis
leads to a similar, but slightly weaker constraint on $g_{\mu\tau}$
for given $m_{Z^\prime}$; this difference presumably results from
inaccuracies of the fast {\tt Delphes 3} simulation of the CMS
detector, as compared to the full simulation based on {\tt Geant
  4}~\cite{Agostinelli:2002hh} employed by the CMS collaboration. For
$Z^\prime$ masses between $10$ and $60$ GeV, this search provides the
strongest bound of all LHC searches.

However, outside this mass range the tightest LHC constraint comes
from other searches. In particular, for $m_{Z^\prime}<10$ GeV the
$4\mu$ search in \cite{Sirunyan:2017lae}, which includes softer muons,
is comparable to or sometimes stronger than
\cite{Sirunyan:2018nnz}. On the other hand, for $m_{Z^\prime}>60$ GeV
the best LHC bound comes from searches for $3\mu$ final states, the
most important ones being \cite{Sirunyan:2017zjc} and, for
$m_{Z^\prime}>100$ GeV, \cite{Sirunyan:2017lae}. Another analysis
\cite{Sirunyan:2018ubx} uses the same selection rules as
\cite{Sirunyan:2017lae} with different categorization, and thus gives
similar results. The main reason for the good performance of the
$3\mu$ searches is that the cross section for the charged current
Drell--Yan process is larger by a factor of $2.5$ to $3$ than that for
the corresponding neutral current process leading to a charged lepton
pair; this relative ordering is not affected much by the $Z^\prime$
boson emitted off the leptons line (see Fig.~1)
\cite{Ma:2001md}. Moreover, the cut efficiency for the most sensitive
$3\mu$ analysis turns out to be a little better.

On the other hand, Fig.~\ref{rst:01} also shows that the LHC bounds
are stronger than existing constraints only in the mass range covered
by the dedicated search \cite{Sirunyan:2018nnz}. Note also that the
upper bounds on the signal cross sections flatten out, or even
slightly increase, at large $Z^\prime$ masses (right frame). This is a
sure sign that the cuts were not optimized for the $L_\mu - L_\tau$
model. For example, the upper bound derived from $3\mu$ final states
in \cite{Sirunyan:2017lae} increases at large $m_{Z^\prime}$ largely because
of a transverse mass cut, which loses efficiency.

So far we have assumed that DM particles cannot be produced in on--shell
$Z^\prime$ decays. If we allow $Z^\prime \rightarrow \phi_{\rm DM} \phi^*_{\rm
  DM}$ decays the branching ratio for $Z^\prime \rightarrow l^+l^-$ decays
will be reduced, leading to reduced $3l$ and $4l$ signals. However,
since we consider a scalar DM particle, even for $q_{\rm DM} = \pm 2$ the
branching ratio for $Z^\prime$ decays into DM particles does not exceed
$25\%$. This would reduce the upper bounds on $g_{\mu\tau}$ derived
from these channels by a factor of at most $\sqrt{12}/4 \simeq 0.86$.

The situation for the $2l$ channel is different. The contribution from
the left Feynman diagram in Fig.~1 to this final state increases with
increasing branching ratio for invisible $Z^\prime$ decays, while that
from the middle diagram decreases. Since for $|q_{\rm DM}| \leq 2$ the
branching ratio for invisible $Z^\prime$ decays is never more than $50\%$,
one might expect the former effect to be dominant; however, the two
diagrams have both different total cross sections and different cut
efficiencies, making a numerical analysis necessary.

\begin{figure}[htb]
\includegraphics[width=0.5\textwidth]{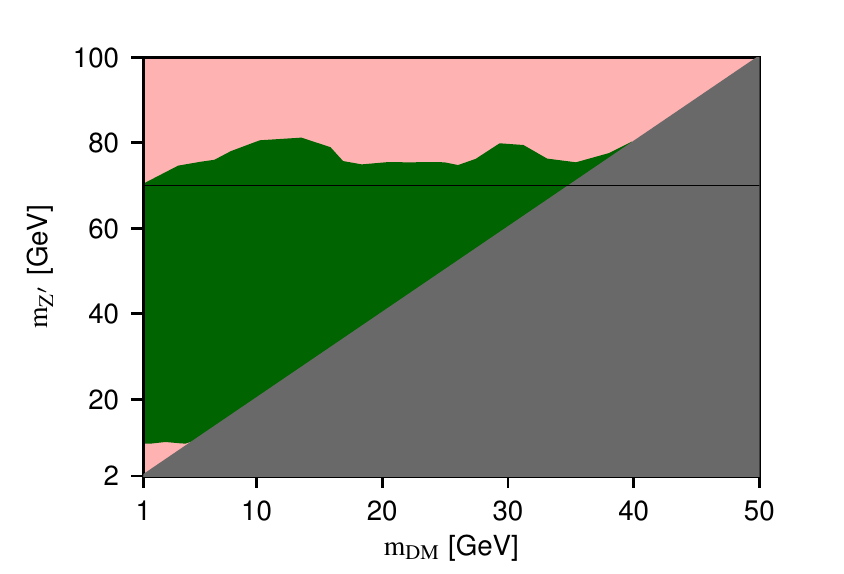}
\includegraphics[width=0.5\textwidth]{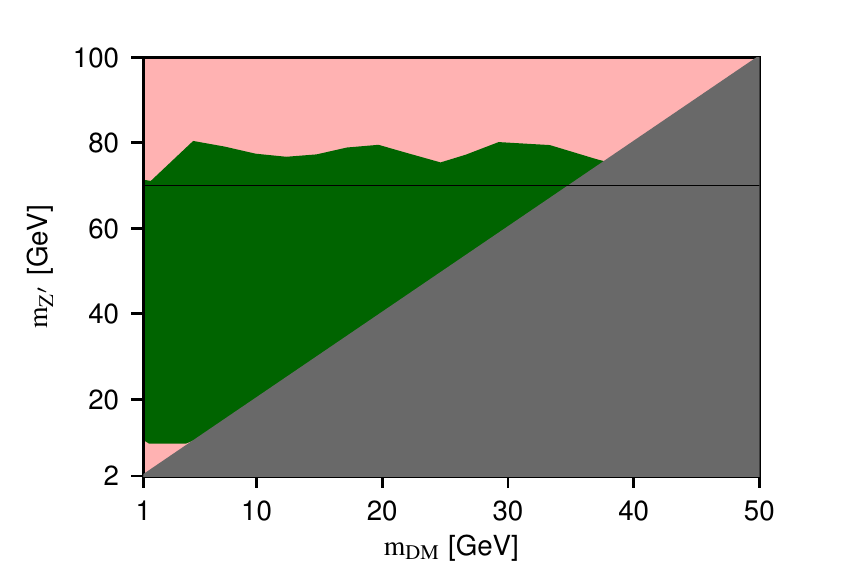}\\
\caption{The effect of $Z^\prime$ decays into dark matter particles on
  the constraint from $2l$ final states, for $g_{\mu\tau}=1$ and
  $q_{\rm DM} = 1$ (left) and $2$ (right). In the gray region below
  the diagonal these decays are kinematically forbidden, i.e. the
  result of Fig.~\ref{rst:01} holds. The green region is excluded by
  analyses of $2\mu$ final states at $\sqrt{s} = 13$ TeV; in the
  absence of $Z^\prime \rightarrow$ DM decays these analyses exclude
  the region between the horizontal lines. The pink region is excluded
  by analyses of $4\mu$ final states, which are only mildly affected
  by $Z^\prime\rightarrow$ DM decays; this includes the entire green
  region.}
\label{rst:02}
\end{figure}

Some results are shown in Fig.~\ref{rst:02}, for $g_{\mu\tau} = 1$ and
$q_{\rm DM} = 1$ (left) and $2$ (right). The green regions are
excluded by our recast of analyses of $2\mu$ final states; the
corresponding exclusion limits in the absence of
$Z^\prime \rightarrow$ DM decays are given by the horizontal black
lines. The fact that the green regions extend beyond the upper
horizontal line shows that allowing
$Z^\prime \rightarrow \phi_{\rm DM} \phi^*_{\rm DM}$ decays increases
the sensitivity of this final state somewhat, the effect being
slightly bigger for $q_{\rm DM} = 2$. The strongest bounds are from three
different analyses of data taken at $\sqrt{s}=13$ TeV
\cite{Sirunyan:2018iwl, Sirunyan:2018egh, Sirunyan:2018lul}, and their
cut efficiencies are indeed quite different for
$pp\rightarrow Z^\prime\nu_l\bar{\nu}_l$ and
$pp\rightarrow Z^\prime l\bar{l}$ processes. However, this entire
region of parameter space is still excluded by analyses of final
states with three or four muons. Therefore LHC data are not sensitive
to the production of dark matter particles in this model.

So far we have considered a complex scalar as dark matter candidate.
However, in on--shell $Z'$ decays the spin of invisible particles cannot
be determined; the only quantity relevant for LHC analyses is the invisible
branching ratio of the $Z'$ boson. For example, we could just as well
consider a Dirac fermion $\chi$ as dark matter candidate. The relevant
partial width would then be given by
\begin{equation} \label{gam_chi}
\Gamma(Z^\prime\rightarrow\bar{\chi}\chi) =
\frac{m_{\rm Z^\prime}}{12\pi}\sqrt{1-4z_{\rm DM}}(g_V^2+g_A^2+2z_{\rm
  DM}(g_V^2-2g_A^2)),
\end{equation}
where the $g_A$ is the axial vector coupling, $g_V$ is the vector
coupling, and $z_{\rm DM} = m^2_\chi / m^2_{Z'}$. For $g_V=0$ and
$g_A=g_{\mu\tau}$, this partial width is the same as that for scalar
DM for $q_{\rm DM}=2$ shown in the right frame of
Fig.~\ref{rst:02}. On the other hand, for $g_A=0$ and
$g_V=g_{\mu\tau}$, eq.(\ref{gam_chi}) predicts a somewhat larger
partial width for sizable mass of the DM particle. However, the
branching ratio for $Z'$ decays into dark matter still remains below
$25\%$, and the constraints from $2\mu + \slashed{E}_T$ searches
remain far weaker than those from analyses of final states with $3$ or
$4$ muons.

\section{Conclusions}
\label{sec:4}

In this study, we recast a large number of LHC analyses, summarized in
Table~\ref{tab:01}, from both the CMS and ATLAS collaborations in the
{\tt CheckMATE} framework in order to constrain the
$U(1)_{L_\mu-L_\tau}$ extension of the SM. Here we focus on the new
$Z^\prime$ gauge boson predicted by this model, whose mass and
coupling are the main free parameters relevant for LHC physics.  We
find that recently published analyses of data taken at $\sqrt{s} = 13$
TeV always have higher sensitivity than LHC data taken at lower
energies. These data exclude $Z^\prime$ masses up to $550$ GeV for
perturbative couplings. We analyzed final states containing two, three
or four charged leptons, where a charged lepton is here defined as a
muon or a hadronically decaying $\tau$ lepton. Final states with only
two charged leptons in principle would have the highest sensitivity to
$Z^\prime$ decays into invisible dark matter particles, but this final
state is always much less sensitive than the $3l$ and $4l$ final
states.  Moreover, replacing a muon by a hadronically decaying $\tau$
lepton always reduced the sensitivity. The final LHC limit is
therefore set by $4\mu$ final states for
$5 \ {\rm GeV} <m_{Z^\prime} <60$ GeV, and by $3\mu$ final states
otherwise. However, except for $10 \ {\rm GeV} < m_{Z^\prime} < 60$
GeV LHC data are still no more sensitive to this model than data taken
at much lower energies, in particular analyses of
$\nu_\mu N \rightarrow \mu^+ \mu^- N$ ``trident'' production by the
CCFR collaboration \cite{Mishra:1991bv}.

Only one analysis we use \cite{Sirunyan:2018nnz}, which looks for $Z
\rightarrow 4\mu$ decays, has been designed specifically for this $Z^\prime$
boson. It is thus very likely that the sensitivity could be enhanced, in
particular for larger $Z^\prime$ masses, by optimizing the cuts, in
particular in $3\mu$ final states which have a larger cross section before
cuts. A further increase of sensitivity might be possible by statistically
combining final states with muons and with hadronically decaying $\tau$
leptons, since the relative normalization of these channels can be predicted
unambiguously in this model; for example, for $m_{Z^\prime} \geq 10$ GeV,
where lepton mass effects are negligible, the branching ratios for $Z^\prime$
decays into $\mu^+\mu^-$ and $\tau^+\tau^-$ are essentially the same.

In this paper we focused on the production of the new $Z^\prime$ gauge boson.
The model also contains a new Higgs boson, which may decay via two real or
virtual $Z^\prime$ bosons into up to four charged leptons. Both the decay of
the $125$ GeV Higgs boson into two of the new Higgs bosons, and the emission
of the new Higgs boson off a $Z^\prime$ boson in one of the diagrams of
Fig.~1, can therefore lead to spectacular final states with up to eight
charged leptons. We leave their investigation to future work.

\FloatBarrier

\Acknowledgements This work was partially supported by the by the
German ministry for scientific research (BMBF). Meng Shi was supported by
the China Scholarship Council (grant no.~201606100045).


\end{document}